\documentclass[twocolumn,showpacs,preprintnumbers,amsmath,amssymb,aps,prb,10pt]{revtex4-1}

\usepackage{graphicx}
\usepackage{subfigure}
\usepackage{bm}

\begin{document}

\title{Magnon spectrum and electron spin resonance in antiferromagnet with large single-ion easy plane anisotropy}

\author{Artemiy S. Sherbakov$^1$}
\email{nanoscienceisart@gmail.com}

\author{Oleg I. Utesov$^{1,2}$}
\email{utiosov@gmail.com}

\affiliation{$^1$National Research Center ``Kurchatov Institute'' B.P.\ Konstantinov Petersburg Nuclear Physics Institute, Gatchina 188300, Russia}
\affiliation{$^2$Department of Physics, Saint Petersburg State University, 198504 St.Petersburg, Russia}

\begin{abstract}

Motivated by recent experiments on quantum magnet NiCl$_{2}$-4SC(NH$_{2}$)$_{2}$ (DTN) and its Br-doped counterpart DTNX we propose a theoretical description of optical magnon branch in the antiferromagnet with large single-ion anisotropy in the magnetically ordered phase. In the framework of the $1/S$ expansion we derive analytical expressions for optical magnon with $\mathbf{k}=0$ energy magnetic field dependence $\Delta(h)$. It is shown that in the linear spin wave approximation $\Delta(h)$ is monotonic without extrema whereas first order in $1/S$ corrections makes it drastically different function with a minimum near the center of magnetically ordered phase. The latter behaviour was observed in ESR experiments. Moreover, we show that $\Delta(h)$ has nontrivial dependence on the system parameters. It solves the discrepancy between inelastic neutron scattering data where the growth of interaction constants in DTNX with small Br concentration was observed and ESR experiments showing almost unchanged $\Delta(h)$ in comparison with pure DTN.

\end{abstract}
\maketitle

\section{Introduction}

Quantum phase transitions have been extensively studied in the last several decades~\cite{sachdev2011}. Among other types of systems magnetic insulators play important role in those investigations (see, e.g., review papers~\cite{Mila, giamarchi2008bose, zheludev2013dirty} and references therein). Magnetic field-induced phase transitions from quantum paramagnet to ordered phases can be usually described theoretically in terms of magnon Bose-Einstein condensation (BEC)~\cite{batyev1984, batyev1985}. Experimentally this phenomenon was observed, for example, in spin-dimer system TlCuCl$_3$~\cite{tlcucl3}.

It is well known that the presence of disorder can significantly influence system properties. Probably, the most famous effect is the localization of elementary excitations~\cite{Anderson1958}. Disordered magnetic systems attracts significant attention now due to the possibility to study the peculiar predictions of the disordered boson physics (so-called ``dirty-boson'') they provide (see Ref.~\cite{zheludev2013dirty} for review). In particular, the existence of a disordered gapless ``Bose glass'' (BG) phase was predicted for dirty bosons~\cite{Fisher}. In systems with quenched disorder this phase always appears between gapped Mott insulator (MI) and gapless superfluid (SF) phases~\cite{theorem}.

Gapped magnets (e.g. spin-dimer systems) are convenient objects to investigate dirty-boson physics if the disorder is realized in magnetic interactions parameters (exchange constants, anisotropy, etc.)~\cite{zheludev2013dirty}. It can be introduced to the system by chemical substitution of non-magnetic ions involved into superexchange interaction~\cite{huvonen2012}.

Dichlorotetrakis-thiourea nickel NiCl$_{2}$-4SC(NH$_{2}$)$_{2}$ (known as DTN) attracts a lot of attention for several reasons. This material is a spin $S=1$ gapped system with large single-ion easy-plane anisotropy. Below $T_{c}\leq 1.2 \, \text{K}$ DTN exhibits antiferromagnetic (AF) ordering in the presence of the external magnetic field directed along the $c$ axis of tetragonal lattice with critical fields $H_{c1}=2.1 \, \text{T}$ and $H_{c2}=12.6 \, \text{T}$~\cite{paduan2004, zapf2005bose}. Remarkably, the critical exponents of this quantum phase transition are those of the BEC universality class with high accuracy~\cite{yin2008, Blinder_2017}. Moreover, DTN is also convenient object for the dirty-boson physics due to a possibility to dope it with bromine (this doped compound is usually referred to as DTNX). In more details, the Cl ions can be replaced by Br ones with minimum changes in the lattice constants and without local changes in the site symmetry\cite{yu2012bose}. Thus, one can use Br substitution to modify interaction constants of Ni spins and introduce bond disorder. At small Br concentrations corresponding couplings can be considered as defects in magnetic subsystem. Their influence on the system properties should be treated either using complicated methods, e.g. bosonic representation of elementary excitations~\cite{sizanovBos} and subsequent accounting for their scattering off defects~\cite{utesov2014}, or introducing some ``effective'', renormalized due to disorder, parameters (see, e.g., Ref.~\cite{povarov2015dynamics}). The latter approach obviously misses some important information, for example disorder-induced quasiparticles damping, and can not catch peculiar behaviour of the system near the phase boundaries such as Bose or Mott glass phases, which were experimentally observed in DTNX in Ref.~\cite{yu2012bose, Orlova_2018}. Nevertheless, effective parameters description is useful in the experimental data interpretation~\cite{povarov2015dynamics, mannig2018} due to its simplicity.

Experimentally, inelastic neutron scattering was used to quantify the effective parameters variation in DTN upon Br substitution~\cite{povarov2015dynamics}. It was shown there that in doped compound with $6\%$ of Br exchange couplings and single-ion anisotropy constant are larger than in the pure DTN, whereas the spectrum gap is lower. This naturally leads to the disappearance of MI phase at higher Br concentrations~\cite{Povarov2017}. In contrast, electron spin resonance (ESR) experiments~\cite{smirnov,smirnov2} show that the ESR spectrum in magnetically-ordered phase stays almost intact upon doping with $7\%$ of Br at $T=0.5 \, \text{K}$. This observation is quite counter-intuitive, because larger system couplings should lead to larger optical magnon energies at $\mathbf{k}=0$ point in the linear spin-wave approximation.

In the present paper we attack this discrepancy analytically. In our analysis we follow Ref.~\cite{sizanov2011antiferromagnet}, but we concentrate on the optical magnons branch rather than low-lying excitations. We use standard Holstein-Primakoff~\cite{holstein} spin operators representation via bosonic ones in order to describe the properties of magnetically-ordered canted antiferromagnetic phase. We find that the ESR spectrum magnetic field dependence calculated using linear spin-wave theory is drastically different from the one which includes the quantum corrections. Moreover, we observe that the ESR spectrum calculated to the first order in $1/S$ expansion describes well the experimental data of Refs.~\cite{smirnov,smirnov2}. Next, taking into account the quantum corrections gives rise to nontrivial spectrum dependence on exchange couplings and single-ion anisotropy constant which allows to solve the described above problem in different experiments data interpretation.

The rest of the paper is organized as follows. In Sec.~\ref{STheor} we discuss general formalism, derive results for the classical spin-wave spectrum, and describe different sources of $1/S$ corrections. Sec.~\ref{SComp} is devoted to comprehensive analysis of the optical magnon in the center of the Brillouin zone energy dependence on different system parameters. The latter are taken close to widely used in literature ones. Sec.~\ref{SSum} contains our conclusion. In the Appendix~\ref{AppA} we present cumbersome part of the Hamiltonian which describes magnon-magnon interaction. Appendix~\ref{AppB} contains some details about important for classical spectrum effective anisotropy constant renormalization.



\section{Theory}
\label{STheor}

DTN is characterised by $I4$ space group and  forms a body-centred tetragonal lattice that may be represented as two interpenetrating  tetragonal subsystems~\cite{Lopez1963}. At $H||c$, the spin dynamics can be described by the Hamiltonian~\cite{zapf2005bose}
\begin{equation}\label{hamiltonian}
  \mathcal{H} = \dfrac{1}{2} \sum_{<i,j>} J_{ij}\textbf{S}_{i} \cdot \textbf{S}_{j}+D\sum_{i}({S_{i}^{z}})^{2}+h\sum_{i}S_{i}^{z}+\mathcal{H}_{int},
\end{equation}
where $J_{ij}$ are exchange constants between the spins and summation is over nearest neighbours, $D$ is the constant of easy-plane single-ion anisotropy, $h=g \mu_{H} H$ --- external magnetic field in energy units.  $\mathcal{H}_{int}$ describes the inter-sublattice interaction which is neglected in the present paper due to its smallness in comparison with the other interactions~\cite{sizanov2011antiferromagnet, Tsyrulin_2013}. Importantly, the following hierarchy of parameters holds in DTN~\cite{zapf2005bose}: $D \gg J_c \gg J_a$, where $J_c$ is the exchange along the $c$ axis of tetragonal lattice, and $J_a$ is in-plane coupling constant. This makes DTN quasi one dimensional system.

Below we consider only magnetically ordered phase at zero temperature in the external magnetic field far from the critical points. The Hamiltonian~\eqref{hamiltonian} is analyzed within the approach similar to developed in Ref.~\cite{sizanov2011antiferromagnet}. However, we focus on the higher in energy optical magnon branch, in particular in $\mathbf{k}=0$ point of the Brillouin zone. This momentum can be directly probed using ESR experiment~\cite{smirnov}. Moreover, we use effective parameters description of the disordered system as it is usually done when DTNX is concerned (see, e.g., Ref.~\cite{povarov2015dynamics}).

\subsection{General formalism}

\begin{figure}
  \centering
  \includegraphics[width=4cm]{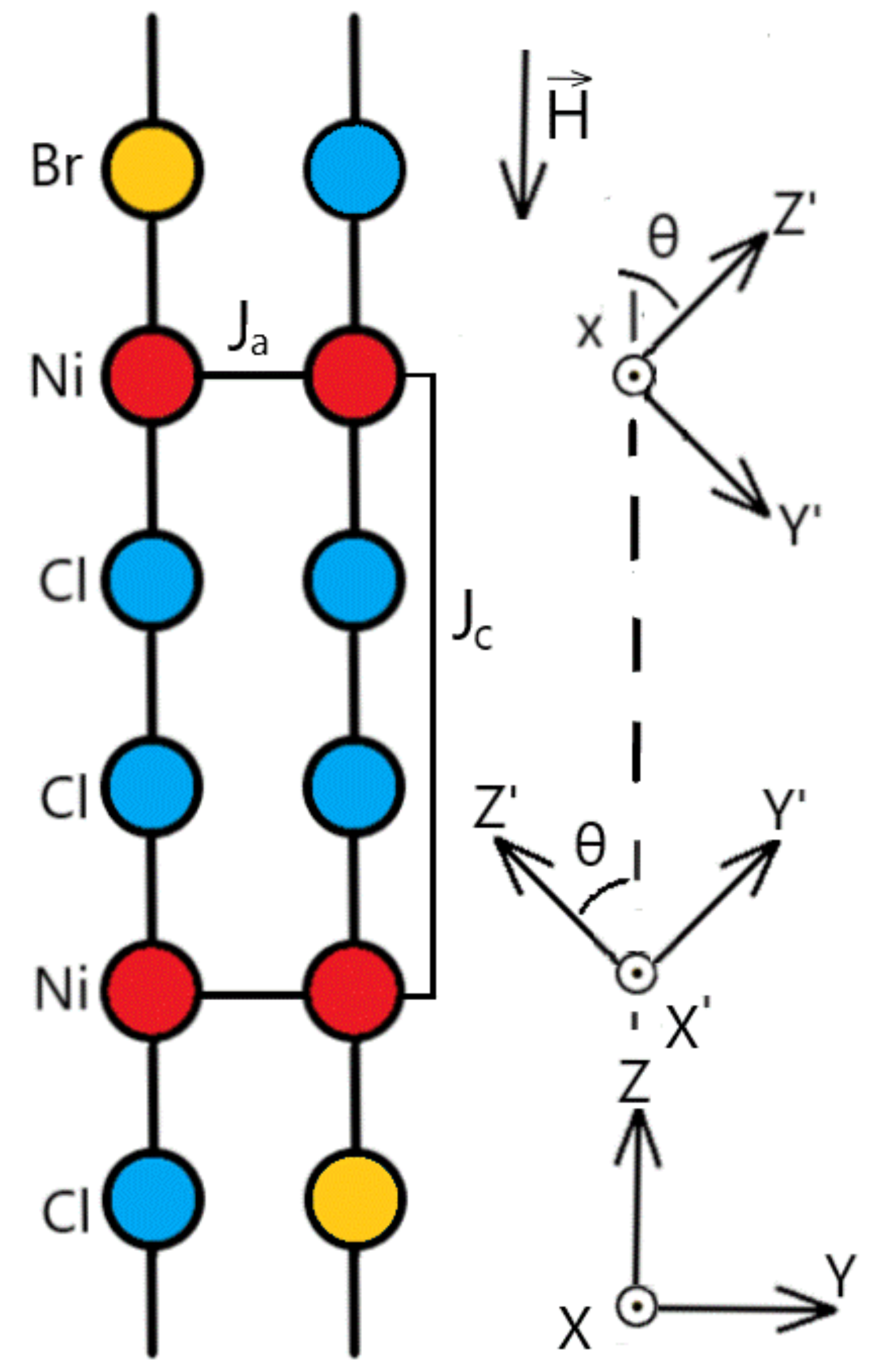}
  \caption{{Canted antiferromagnetic ordering in the external magnetic field. In our analysis we introduce local quantization axes $z^\prime$ for spins in each of two magnetic sublattices, which orientation is shown by arrows.}
\label{spins}}
\end{figure}

Since our goal is to describe the properties of the canted AF phase it is convenient to introduce~\cite{sizanov2011antiferromagnet} a local coordinate frame $(x^\prime,y^\prime,z^\prime)$ on each site. The mean spin value on each site is assumed to be parallel to $z^\prime$ axis (see Fig.~\ref{spins}). Spin components in the local coordinate frame are expressed via those in laboratory coordinate frame as follows:
\begin{equation}\label{locspin}
  \begin{gathered}
  S_{i}^{x}=S_{i}^{x^\prime}, \\
  S_{i}^{y}= S_{i}^{y^\prime}\cos{\theta}+S_{i}^{z^\prime}\exp(i\textbf{k}_{0}\textbf{R}_{i})\sin{\theta},\\
  S_{i}^{z}= S_{i}^{z^\prime}\cos{\theta}-S_{i}^{y^\prime}\exp(i\textbf{k}_{0}\textbf{R}_{i})\sin{\theta},
\end{gathered}
\end{equation}
where $\textbf{k}_{0}=(\pi,\pi,\pi)$ is the AF vector, and the imaginary exponents describe the Neel ordering in the XY-plane. Henceforth we put all the distances between neighbouring spins in the lattice to be equal to unity.

In the subsequent calculations of the magnon spectrum we take into account the first order in $1/S$ contributions. So, we use the Holstein-Primakoff~\cite{holstein} representation of spins operators via bosonic ones in the local coordinate frame in the following from:
\begin{equation}\label{spinresp}
\begin{gathered}
S_{i}^{x^\prime}+iS_{i}^{y^\prime}=\sqrt{2S}a^{\dagger}_{i}\sqrt{1-\dfrac{a^{\dagger}_{i}a_{i}}{2S}}\approx\sqrt{2S}a^{\dagger}_{i}\left(1-\dfrac{a^{\dagger}_{i}a_{i}}{4S}\right),\\
S_{i}^{x^\prime}-iS_{i}^{y^\prime}=\sqrt{2S}\sqrt{1-\dfrac{a^{\dagger}_{i}a_{i}}{2S}}a_{i}\approx\sqrt{2S}\left(1-\dfrac{a^{\dagger}_{i}a_{i}}{4S}\right)a_{i},\\
S_{i}^{z^\prime}=-S+a^{\dagger}_{i}a_{i}.
\end{gathered}
\end{equation}
Using Eqs.~\eqref{locspin} and~\eqref{spinresp} and producing the Fourier transform of $a_i$ and $a^\dagger_i$ operators ($N$ being a number of magnetic ions)
\begin{eqnarray} \label{Fourier}
  a_i = \frac{1}{\sqrt{N}} \sum_\mathbf{k} a_\mathbf{k} e^{i \mathbf{k} \cdot \mathbf{R}_i},\\ \nonumber
  a^\dagger_i = \frac{1}{\sqrt{N}}\sum_\mathbf{k} a^\dagger_\mathbf{k} e^{- i \mathbf{k} \cdot \mathbf{R}_i},
\end{eqnarray}
we obtain the Hamiltonian~\eqref{hamiltonian} to the first order in $1/S$ as a sum of five terms,
\begin{equation}
  \mathcal{H}=\sum_{i=0}^{4} \mathcal{H}_{i},
\end{equation}
where
\begin{equation} \label{E0}
\dfrac{1}{N} \mathcal{H}_{0}=S^{2}\left[(J_{0}+\widetilde{D})\cos^{2}\theta-\dfrac{h\cos\theta}{S}-\tilde{D}+D\right],
\end{equation}
\begin{equation} \label{Hlinear}
\dfrac{1}{\sqrt{N}}\mathcal{H}_{1}=i(a_{\mathbf{k}_{0}}-a_{\mathbf{k}_{0}}^{\dagger})\sqrt{\frac{S}{2}}[2S(J_{0}+\widetilde{D})\cos\theta-h]\sin\theta,
\end{equation}
\begin{eqnarray} \label{Hbilinear}
 \mathcal{H}_{2}=&&\sum_{\mathbf{k}}a_{\mathbf{k}}^{\dagger}a_{\mathbf{k}} \left[E_{\mathbf{k}}-\dfrac{D}{2}(1-3\cos^{2}\theta) \right]+\nonumber
 \\+\frac{1}{2}&&\sum_{\mathbf{k}}(a_{\mathbf{k}}a_{-\mathbf{k}}+a_{\mathbf{k}}^{\dagger}a_{-\mathbf{k}}^{\dagger})\left[B_{\mathbf{k}}+\dfrac{D}{2} \sin^{2}\theta \right].
\end{eqnarray}
Here
\begin{eqnarray} 
E_{\mathbf{k}}=&&S(J_{0}+J_{\mathbf{k}})\cos^{2}\theta+S(J_{0}+D)(1-3\cos^{2}\theta)\nonumber\\ \label{E0} +&& h\cos\theta \\ \label{B0}
B_{\mathbf{k}}=&&S(J_{\mathbf{k}}-D)\sin^{2}\theta, \\ 
J_\mathbf{k} = && 2 \left[ J_c \cos{k_z} + J_a (\cos{k_x}+\cos{k_y}) \right],
\end{eqnarray}
and cumbersome expressions for $\mathcal{H}_3$ and $\mathcal{H}_4$ (which contain products of three and four bosonic operators, respectively) are presented in Appendix~\ref{AppA}.

Considering a single-ion anisotropy, one must take into account that for $S=1/2$ this interaction leads only to a constant correction to the energy of the ground state. Therefore, using $1/S$ expansion all terms containing $D$ must disappear if $S=1/2$. This condition is manifested in the fact that in all orders of the perturbation theory constant $D$ will enter as renormalized\cite{Kaganov_1987}(see also Appendix~\ref{AppB} for some simple arguments concerning linear spin-wave theory). Thus, the effective single-ion anisotropy constant reads:
\begin{equation} \label{TildeD}
\widetilde{D}=D\left(1-\dfrac{1}{2S}\right).
\end{equation}
It can be directly used when the spectrum of linear theory is considered (see Eq.~\eqref{Hbilinear} where there are $1/S$ corrections to Eqs.\eqref{E0} and~\eqref{B0} formally substituting $D$ by $\tilde{D}$). However, $1/S$ corrections should be treated in a more accurate way (see below).

In order to calculate magnon spectrum to the first order in $1/S$ it is convenient to introduce the following Green's functions:
\begin{equation}
  G_{k}= \langle a_{\textbf{k}}, a_{\textbf{k}}^{\dagger} \rangle_{\omega}; \, F^\dagger_{k}=\langle a_{-\textbf{k}}^{\dagger}, a_{\textbf{k}}^{\dagger} \rangle_{\omega},
\end{equation}
where $k=(\omega,\textbf{k})$. These functions obey the following system of Dyson's equations
\begin{eqnarray} \label{Dyson}
  G_{k}=G^{0}_{k}+G^{0}_{k} \, \Sigma_{k}\, G_{k}+G^{0}_{k}(B_{\textbf{k}}+\Pi_{k})F_{k}, &&\\ \nonumber
  F_{k}=G^{0}_{-k}(B_{\textbf{k}}+\Pi^\dagger_{k})G_{k}+G^{0}_{-k} \, \Sigma_{-k} \, F_{k},
\end{eqnarray}
where $\Sigma$ and $\Pi$ are normal and anomalous self-energy parts, respectively, $G^{0}_{k}=(\omega - E_\mathbf{k} + i\delta)^{-1}$ is bare Green's function, $B_{\textbf{k}}$  is the anomalous self-energy part of zeroth order. Evidently, due to the hermitianity of the Hamiltonian~\eqref{hamiltonian} $\Pi^\dagger_k = \Pi_k$.

After solving the system of equations~\eqref{Dyson} we obtain the following expression for normal and anomalous Green's functions
\begin{eqnarray} \label{Green}
G_{k}=\dfrac{\omega+E_{\textbf{k}}+\Sigma_{-k}}{\omega^{2}-{\varepsilon^{(0)}_{\mathbf{k}}}^{2}-\Omega_{k}}; \\ F_{k}=-\dfrac{B_{\textbf{k}}+\Pi_{k}}{\omega^{2}-{\varepsilon^{(0)}_{\mathbf{k}}}^{2}-\Omega_{k}},
\end{eqnarray}
where the bare magnon spectrum $\varepsilon^{(0)}_{k}$ reads
\begin{equation}
\varepsilon^{(0)}_{\mathbf{k}}=\sqrt{{E^2_{\mathbf{k}}}-{B^2_{\mathbf{k}}}},
\label{clpecrtum}
\end{equation}
and $\Omega_{k}$ describes the renormalization of magnon spectrum. Since we are interested in quantum corrections in the first order in $1/S$ we use expression
\begin{equation} \label{Omega}
\begin{gathered}
\Omega_{k}=\omega(\Sigma_{k}-\Sigma_{-k})
+E_{\textbf{k}}(\Sigma_{k}+\Sigma_{-k})-2B_{\textbf{k}}\Pi_{k}
\end{gathered}
\end{equation}
below.

\subsection{Classical picture}

In this Subsec. we briefly discuss well-known results of the linear spin-wave approximation.

First of all, we should find a relation between canting angle $\theta$ and the external magnetic field $h$ in zeroth order in $1/S$. This can be done either by minimizing the classical energy $\mathcal{H}_{0}$~\eqref{E0} or by making the coefficient in $\mathcal{H}_{1}$~\eqref{Hlinear} to be equal to zero. As a result neglecting small in $1/S$ terms we have
\begin{equation} \label{theta}
 \cos{\theta} = \left\{
  \begin{array}{ll}
  h/h^{(0)}_{c}  &, \, h < h^{(0)}_{c} \textrm{,}\\
  1 & , \, h \geq h^{(0)}_{c},
  \end{array} \right.
\end{equation}
where $h^{(0)}_{c}=2S(J_{0}+D)$ is the classical saturation field.

\begin{figure}
  \centering
  \includegraphics[width=6cm]{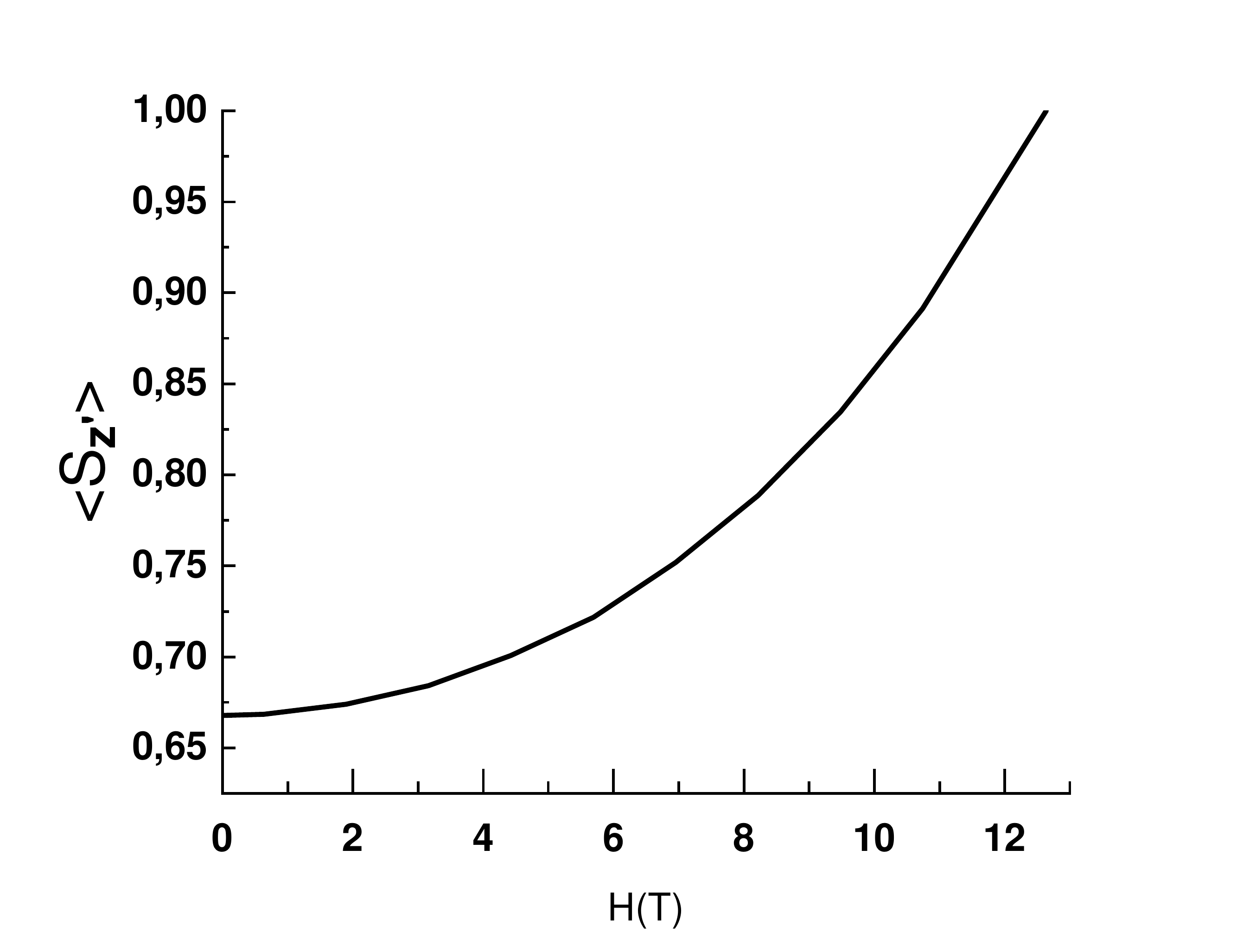}
  \caption{Average value of the ordered spin component calculated in the linear spin-wave approximation using Eq.~\eqref{meanspin} at $T=0$ and the parameters $D=8.9 \, \text{K}, J_{c}=2.2 \, \text{K}, J_{a}=0.18 \, \text{K}$ of Ref.\cite{Zvyagin2007}
\label{figspin}}
\end{figure}

Next, using Eq.~\eqref{theta} we can write
\begin{eqnarray}\label{Selfparts0}
  E_{\mathbf{k}}=S[(J_{0}+J_\mathbf{k})\cos^{2}{\theta}+(J_{0}+D)\sin^{2}{\theta}],&&\\ \nonumber
  B_{\mathbf{k}}=S (J_\mathbf{k}-D)\sin^{2}{\theta},
\end{eqnarray}
and obtain the classical spectrum~\eqref{clpecrtum} in the following well-known form
\begin{eqnarray} \label{clspectrum2}
\varepsilon^{(0)}_{\mathbf{k}}=S\sqrt{(J_{0}+J_{\mathbf{k}})(J_{0}+J_{\mathbf{k}}\cos{2\theta}+2 D\sin^{2}\theta)}.
\end{eqnarray}
Finally, it is easy to write down expression for optical magnon with $\mathbf{k}=0$ energy in the magnetic field,
\begin{equation}
  \Delta^{(0)}(h) \equiv \varepsilon^{(0)}_{0}=2S \sqrt{J_{0}(J_{0}\cos^{2}\theta+D\sin^{2}\theta)}.
  \label{delta0}
\end{equation}
Evidently, this quantity has very simple dependence on the interactions constants $J$ and $D$. At given magnetic field parameters increasing leads to growth of $\Delta^{(0)}(h)$ and their decreasing leads to lower values of $\Delta^{(0)}(h)$.

As it was already noticed usually $\widetilde{D}$ (see Eq.~\eqref{TildeD}) is used in the linear spin-wave approximation instead of $D$, which is equivalent to certain resummation of the perturbation theory rows. Formally, contributions to the spectrum from magnon-magnon interaction are of the same order in $1/S$ as terms included by using $\tilde{D}$ instead of $D$. However, if we for a while forget about non-linear corrections we can define renormalized saturation field
\begin{equation}\label{Hsat}
  h_{c}=2S(J_{0}+\tilde{D}),
\end{equation}
which should be plugged into canting angle definition $\cos{\theta}= h/h_c$. Importantly this field is not a subject of further corrections (see below) and is an exact one.

Next, we can include small in $1/S$ terms from Eq.~\eqref{Hbilinear} into classical spin-wave spectrum. Along with canting angle renormalization with the use of Eq.~\eqref{Hsat} we arrive to the classical spectrum~\eqref{clspectrum2} where $D$ is substituted by $\tilde{D}$. Evidently, this can not change the conclusion about $\Delta^{(0)}(h)$ dependence on the interactions parameters in the linear spin-wave approximation.

We also point out that our approach can not describe the properties of DTN at low magnetic fields, where in fact the system is in a gapped phase without magnetic order. To illustrate this we calculate  average spin value, which reads
\begin{equation} \label{meanspin}
  \langle S_{z^\prime} \rangle = S + \dfrac{1}{2}-\int\dfrac{d\mathbf{k}}{(2\pi)^{3}}\dfrac{E_{\mathbf{k}}}{2\varepsilon^{(0)}_{\mathbf{k}}}.
\end{equation}
Its magnetic field dependence is shown in Fig.~\ref{figspin}. One can see that the average value of ordered spin is nonzero even at zero magnetic field, however the fluctuations-induced correction is quite large, being approximately $1/3$. Moreover, taking into account $1/S$ corrections leads to further diminishing of $\langle S_{z^\prime} \rangle$ at low magnetic fields. This also pinpoints the importance of quantum corrections in the considered system.

\subsection{The first order corrections}

Here we discuss the renormalized magnon spectrum due to the first order in $1/S$ quantum corrections.

In this order of the perturbation approach we derive analytical expression for Eq.~\eqref{Omega} which leads to the following magnon spectrum:
\begin{equation}
  \varepsilon^{(1)}_{\textbf{k}}=\varepsilon^{(0)}_{\mathbf{k}}+\dfrac{\Sigma_{k}-\Sigma_{-k}}{2}+\dfrac{E_{\textbf{k}}(\Sigma_{k}+\Sigma_{-k})-2B_{\textbf{k}}\Pi_{k}}{2\varepsilon^{(0)}_{\mathbf{k}}},
\end{equation}
where instead of $\omega$ in all the self-energy parts one should use $\varepsilon^{(0)}_{\mathbf{k}}$ given by Eq.~\eqref{clspectrum2} without $D \rightarrow \tilde{D}$ substitution.

\begin{figure}
\includegraphics[width=1.0\linewidth]{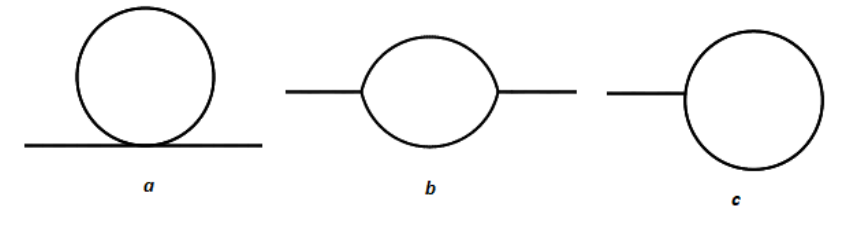}
\caption{Diagrams illustrating contributions to the magnon spectrum in the first order in $1/S$. Solid lines stands for both normal and anomalous Green's functions $G_k$ and $F_k$ (see Eqs.~\eqref{Green}) of zeroth order. (a) Hartree-Fock type diagrams due to the $\mathcal{H}_4$ part of the Hamiltonian~\eqref{H4}. (b) Loop type diagrams originating from $\mathcal{H}_3$. (c) Diagrams which provide quantum correction to the linear term in the Hamiltonian~\eqref{Hlinear}.}
\label{diagrams}
\end{figure}

There are several types of corrections to the bare magnon spectrum $\varepsilon^{(0)}_{\mathbf{k}}$. First of all, one should take into account terms in the bilinear part of Hamiltonian~\eqref{Hbilinear} which do not contain $S$. Next, there are three contributions to the magnon self-energy parts which are given by the following diagrams: (i) Hartree-Fock type corrections shown in Fig.~\ref{diagrams}(a) which stems from perturbation Hamiltonian $\mathcal{H}_4$~\eqref{H4}, (ii) loop diagrams (Fig.~\ref{diagrams}(b)) originating from $\mathcal{H}_3$~\eqref{H3}, (iii) diagrams shown in Fig.~\ref{diagrams}(c) which lead to $1/S$ correction to the linear term in the Hamiltonian~\eqref{Hlinear}.

While we do not present cumbersome expressions for contributions (i) and (ii), we briefly discuss (iii). It results in the following correction to the Hamiltonian
\begin{equation}
\frac{1}{\sqrt{N}}H^{(1)}_{1}=i\dfrac{\sqrt{S}}{2}\sin\theta\cos\theta(a_{\mathbf{k}_{0}}-a_{\mathbf{k}_{0}}^{+})A(h),
\label{H_13S}
\end{equation}
where
\begin{equation}
A(h)=-\frac{1}{N}\sum_{\mathbf{k}}\dfrac{(E_{\mathbf{k}}-\varepsilon^{(0)}_{\mathbf{k}})(V_{\mathbf{k}_{0}}+V_{\mathbf{k}})+V_{\mathbf{k}}B_{\mathbf{k}}}{8\varepsilon^{(0)}_{\mathbf{k}}},
\end{equation}
and $V_\mathbf{k}$ is given by Eq.~\eqref{Vk}. Correction~\eqref{H_13S} leads to renormalization of the classical canting angle $\theta$; new angle $\tilde{\theta}$ is defined by
\begin{equation}
\cos\tilde{\theta}=\cos\theta\left[1-A(h)/h_{c}\right].
\label{rcos}
\end{equation}
This quantity should be used when taking into account $1/S$ corrections to bare self-energies $E_\mathbf{k}$ and $B_\mathbf{k}$. It can be shown that the quantum correction to the canting angle vanishes at $h=h_{c}$, thus it does not affect the second critical field value.

Taking all the $1/S$ contributions into account we obtain cumbersome analytical expression for $\varepsilon^{(1)}_{\textbf{k}}$, and subsequently for $\Delta^{(1)}(h) \equiv \varepsilon^{(1)}_0$, the quantity of prime importance for the present study. Its behaviour in DTN and DTNX is analyzed in details in the next Sec. Here we would like to point out once again that the result for $\Delta^{(1)}(h)$ is reliable only in the range of magnetic fields in the ordered phase not very close to the critical fields $H_{c1}$ and $H_{c2}$.




\section{Comparison with experiment}
\label{SComp}

We begin with comparison of the classical magnon energy at $\mathbf{k}=0$ $\Delta^{(0)}(h)$ given by Eq.~\eqref{delta0}, the one which includes quantum corrections $\Delta^{(1)}(h)$, and observed experimentally in pure DTN in Ref.~\cite{smirnov} at $T=0.5 \, \text{K}$ ESR spectrum. Notice, that our calculations are made at zero temperature. However, we believe that they can be used for the experimental data analysis since $T=0.5 \, \text{K}$ is significantly smaller than $T_c$ when magnetic field is not very close to the critical ones.

\begin{figure}[t]
  \centering
  \includegraphics[width=7cm]{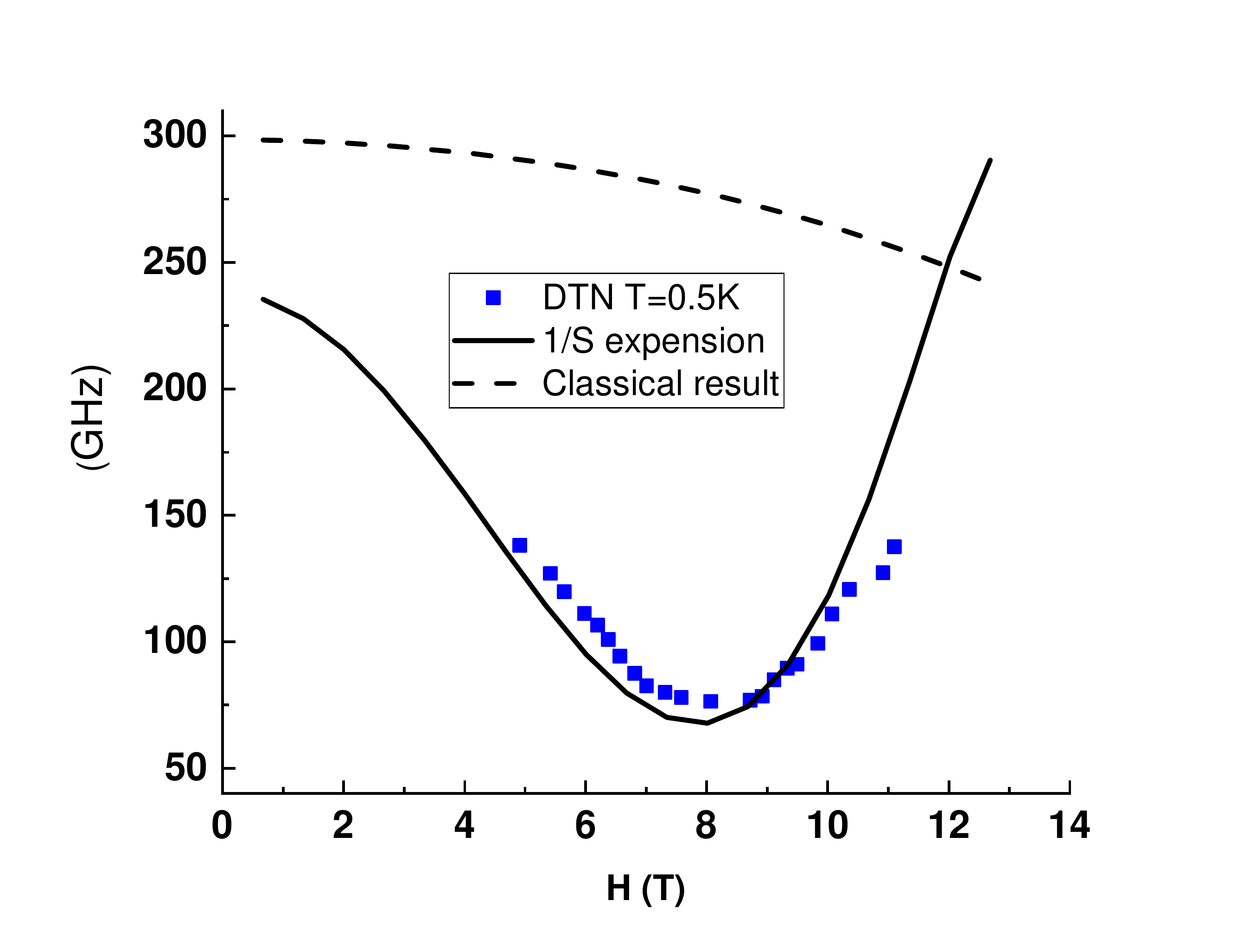}
  \caption{Comparison of the magnetic field dependence of magnon with $\mathbf{k}=0$ energy $\Delta(h)$ calculated using the linear spin-wave approximation Eq.~\eqref{delta0} (dashed line), its counterpart $\Delta^{(1)}(h)$, which includes first order in $1/S$ corrections (solid line), and the ESR experiment data from Ref.~\cite{smirnov}(square dots). Here parameters $D=9.1 \, \text{K}, J_{c}=2.5 \, \text{K}, J_{a}=0.15 \, \text{K}$ were used.
\label{comparison}}
\end{figure}

In Fig.~\ref{comparison} we show theoretically calculated (using particular set of parameters $D=9.1 \, \text{K}, J_{c}=2.5 \, \text{K}, J_{a}=0.15 \, \text{K}$) magnetic field dependencies of $\Delta^{(0)}(h)$ and $\Delta^{(1)}(h)$. One can see that the classical curve is monotonic and has no extrema, whereas its counterpart which includes $1/S$ corrections is non-monotonic function with the minimum near the center of magnetically ordered phase. Importantly, the latter has a very good quantitative agreement with the experimental data of Ref.~\cite{smirnov} ($100 \, \text{GHz} \approx 4.78 \, \text{K}$). Furthermore, this drastic qualitative difference between two theoretically obtained curves stays intact when various sets of parameters relevant to DTN and DTNX are considered (taken, e.g., from Refs.~\cite{sizanovBos,Tsyrulin_2013,povarov2015dynamics}). It shows that quantum corrections are very important in DTN properties description, which makes this compound very different from standard ordered magnets.

\begin{figure}[h]
  \centering
\begin{minipage}[h]{0.45\linewidth}
\center{\includegraphics[width=1.2\linewidth]{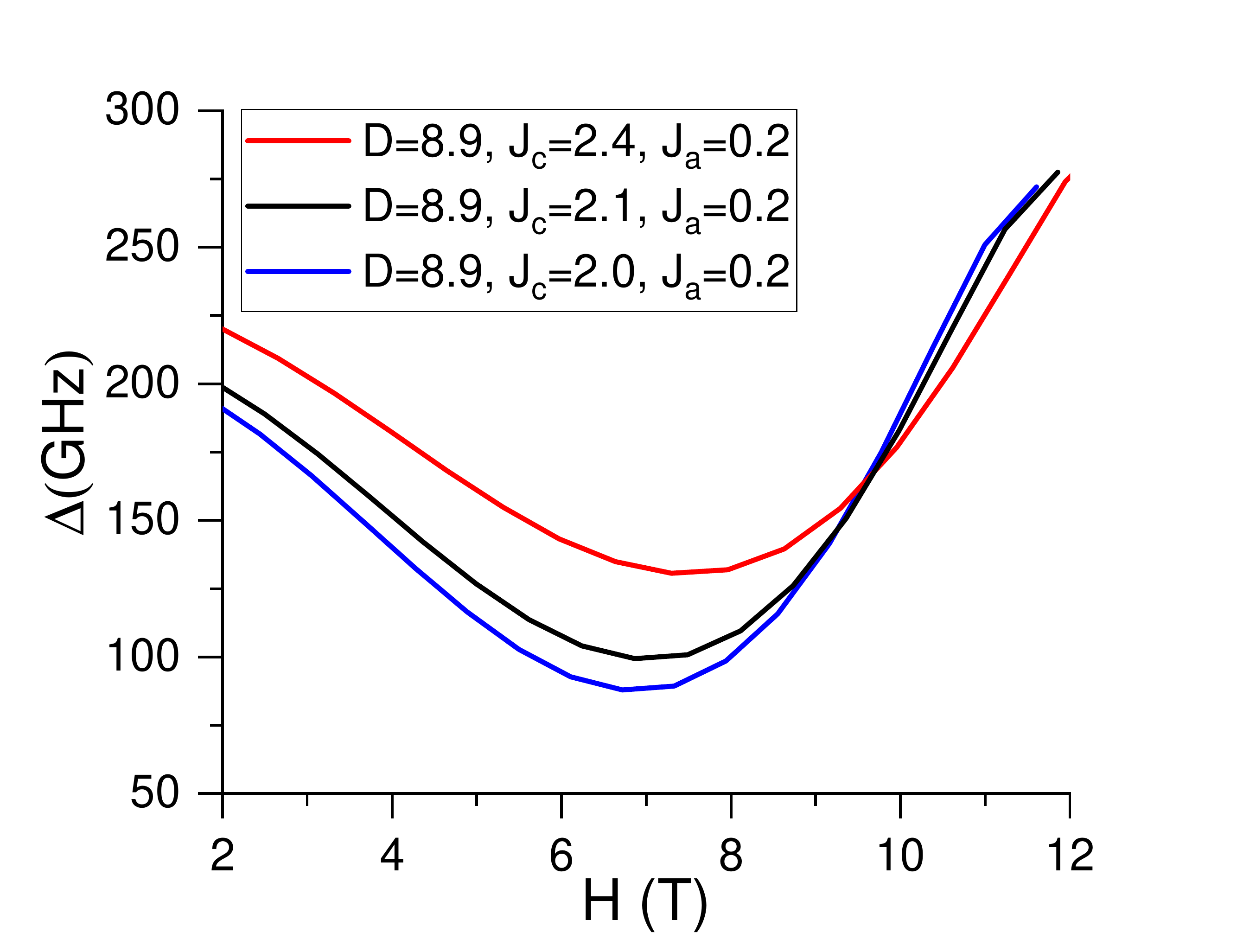}} a \\
\end{minipage}
\hfill
\begin{minipage}[h]{0.45\linewidth}
\center{\includegraphics[width=1.2\linewidth]{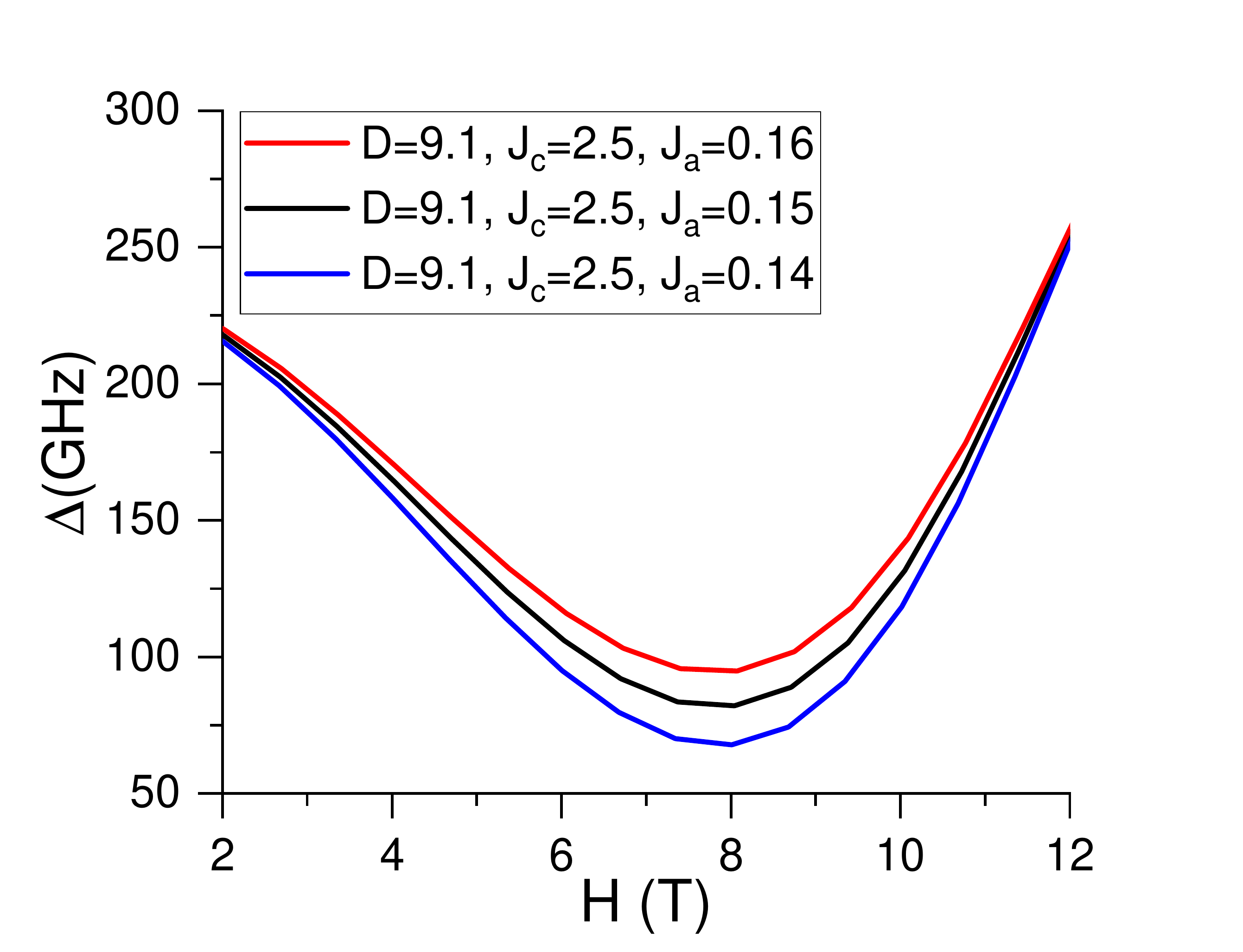}} b \\
\end{minipage}
\hfill
\begin{minipage}[h]{0.45\linewidth}
\center{\includegraphics[width=1.2\linewidth]{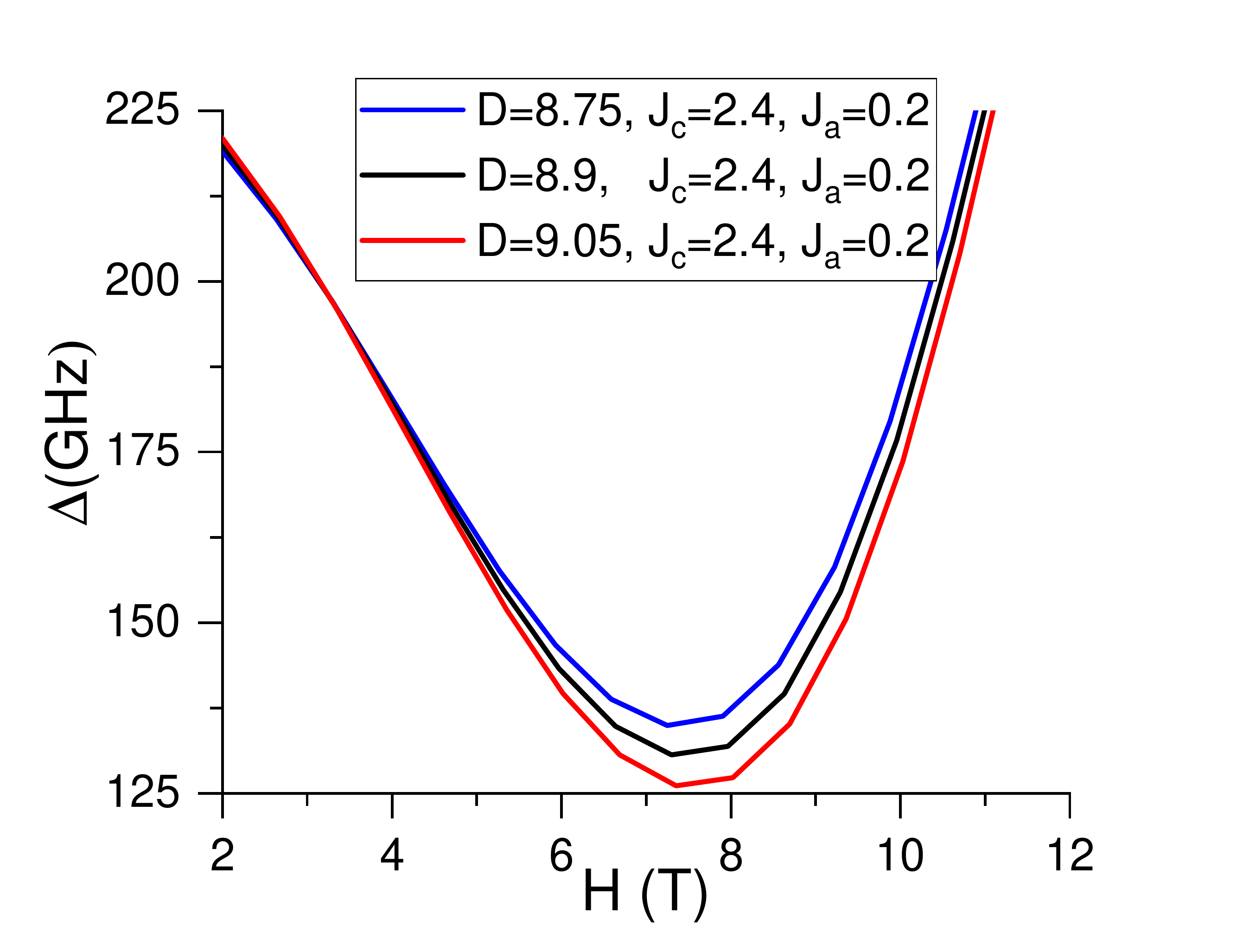}} c \\
\end{minipage}
\hfill
\begin{minipage}[h]{0.45\linewidth}
\center{\includegraphics[width=1.2\linewidth]{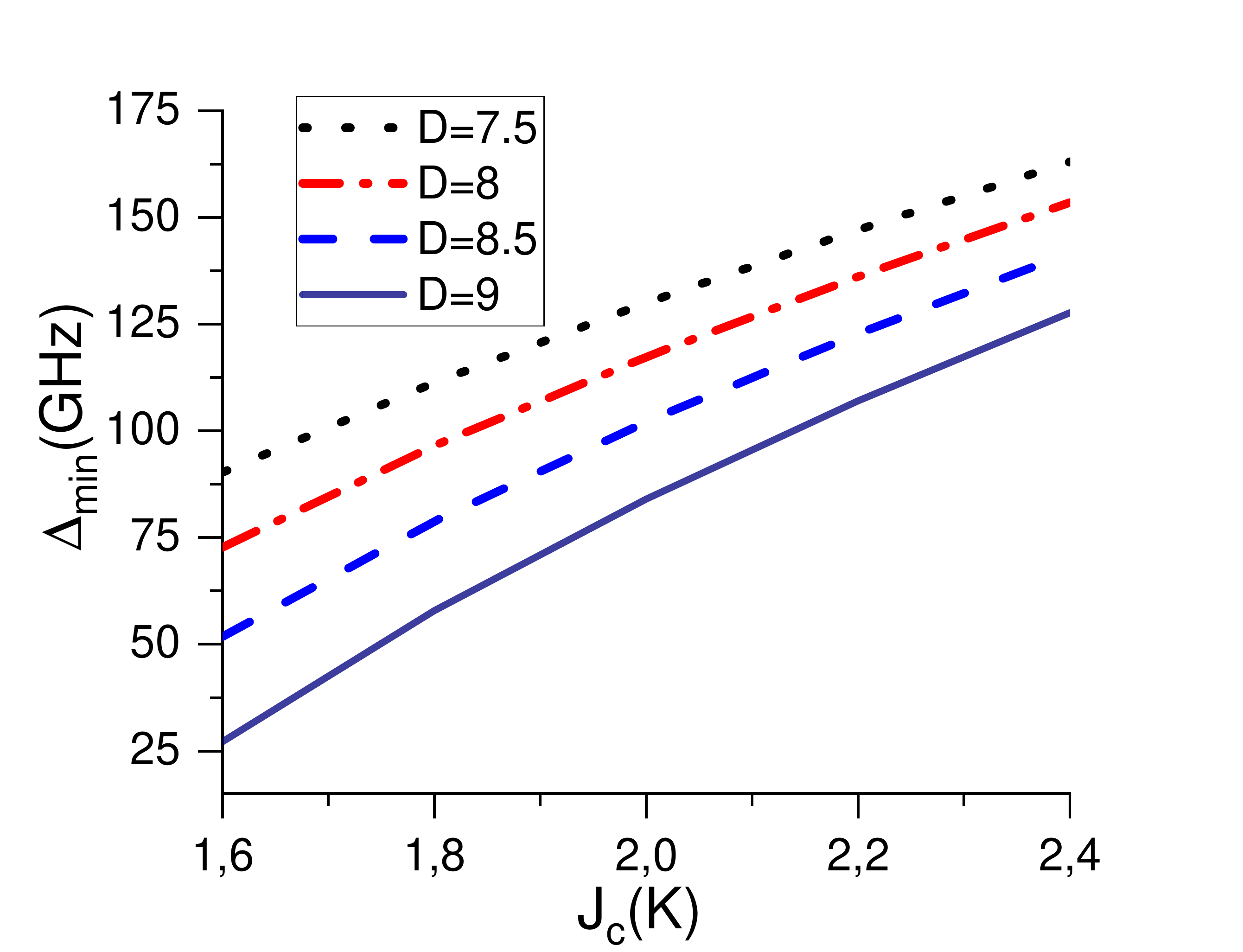}} d \\
\end{minipage}
\caption{Examples of  $\Delta^{(1)}(h)$ for different parameters values (a,b,c). (d) The dependence of the minimal value of $\Delta^{(1)}(h)$ as a function of magnetic field --- $\Delta_{min}$ on the system parameters $D, J_{c}$ at fixed $J_{a}=0.2 \, \text{K}$. Evidently, growth of $D$ and $J_{c}$ has opposite effect on $\Delta_{min}$ (see text).
\label{fig1}}
\end{figure}

It was already pointed out above that $\Delta^{(0)}(h)$ behaviour under system parameters variation is very simple. For example if $D$ and $J_c$ increase $\Delta^{(0)}(h)$ also increases. In contrast, $\Delta^{(1)}(h)$ dependence on the parameters is much more complicated. Generally, anisotropy constant and exchange couplings variation have opposite effect on this quantity. In more details, influence of the system parameters variation on $\Delta^{(1)}(h)$ can be summarized the following way:
\begin{itemize}
  \item If $D$ and $J_{a}$ are fixed and $J_{c}$ increases, $\Delta^{(1)}(h)$ increases;
  \item If $D$ and $J_{c}$ are fixed and $J_{a}$ increases, $\Delta^{(1)}(h)$ increases;
  \item If $J_{a}$ and $J_{c}$ are fixed and $D$ increases, $\Delta^{(1)}(h)$ decreases.
\end{itemize}
We illustrate some of this statements in Fig.~\ref{fig1}, where we present  $\Delta^{(1)}(h)$ for different parameters values and   analyse minimal value of $\Delta^{(1)}(h)$ as function of magnetic field --- $\Delta_{min}$. One can see that at fixed $J_a = 0.2 \, \text{K}$ $\Delta_{min}$ increases with $J_c$ increase and decreases with $D$ increase.

Finally, we want to stress that our analysis explains the fact that ESR spectrum of DTN can remain almost unchanged~\cite{smirnov2} when small concentrations of Br are introduced, while according to the neutron experiment exchange couplings and single-ion anisotropy constant increase with the increase of Br concentration~\cite{povarov2015dynamics}. From the point of view of $\Delta^{(1)}(h)$ in our theory different system parameters variation can compensate each other, which can not occur in the linear spin-wave approximation.

\section{Summary}
\label{SSum}

Motivated by the recent experiments on the Br doped DTN (so called DTNX)~\cite{povarov2015dynamics,smirnov2} we discuss theoretically the spin-wave spectrum of antiferromagnet with large single-ion easy-plane anisotropy in the magnetic field induced ordered phase. In particular, we obtain analytical expression for energy of the optical magnon $\Delta(h)$ at the center of the Brillouin zone which can be measured in ESR experiment. We showed that the dependence $\Delta(h)$ in the linear spin-wave approximation is drastically different from the one which includes quantum corrections of the order $1/S$. The latter are responsible for nonmonotonic dependence of $\Delta(h)$ on the external magnetic field which was observed experimentally in both DTN~\cite{smirnov} and DTNX~\cite{smirnov2}. Furthermore, according to our analysis quantum corrections give rise to nontrivial $\Delta(h)$ dependence on the system parameters in contrast to the simple classical behaviour. We show that while effective exchange couplings and single-ion anisotropy increase in DTNX as compared to DTN at small Br concentrations~\cite{povarov2015dynamics}, $\Delta(h)$ can stay almost unchanged, as it was experimentally observed in Ref.~\cite{smirnov2}.


\begin{acknowledgments}

We are grateful to A. I. Smirnov and A. V. Syromyatnikov for stimulating discussions. The reported study was supported by the Foundation for the Advancement of Theoretical Physics and Mathematics ``BASIS''

\end{acknowledgments}

\appendix

\section{Expressions for $\mathcal{H}_{3}$ and $\mathcal{H}_{4}$}
\label{AppA}

Here we present cumbersome expressions for $\mathcal{H}_{3}$ and $\mathcal{H}_{4}$ terms in the Hamiltonian which include three and four bosonic operators, respectively. The former reads:
\begin{eqnarray} \label{H3}
\sqrt{N} \mathcal{H}_{3}&=&i\dfrac{\sqrt{S}}{4\sqrt{2}}\sin\theta\cos\theta \\ \nonumber &&\sum_{123}(a_{1}^{+}a_{2}^{+}a_{-3}-a_{-3}^{+}a_{2}a_{1})\dfrac{V_{1}+V_{2}}{2}
\end{eqnarray}
where indexes $1,2,3$ stand for different momenta $\textbf{k}_{1}, \textbf{k}_{2},\textbf{k}_{3}$, the conservation law $\textbf{k}_{1}+\textbf{k}_{2}+\textbf{k}_{3}=\textbf{k}_{0}$ is implied, and we denote
\begin{equation}\label{Vk}
V_{\mathbf{k}}=\left(2J_{0}-8J_{\mathbf{k}}+10D-\dfrac{h}{S\cos\theta}\right).
\end{equation}

Expression for $\mathcal{H}_{4}$ has the following form:
\begin{eqnarray}
&& N \mathcal{H}_{4}=\sum_{1234}a_{-1}^{+}a_{-2}^{+}a_{3}a_{4} \Biggl[\dfrac{1-2\sin^{2}\theta}{2}J_{4-1} \nonumber \\ \label{H4} && -\dfrac{\cos^{2}\theta}{4}(J_{1}+J_{4})+D\left(1-\dfrac{3}{2}\sin^{2}\theta\right)\Bigg] \\ &&+(a_{1}^{+}a_{2}^{+}a_{3}^{+}a_{-4}+a_{-4}^{+}a_{3}a_{2}a_{1})\left[\dfrac{D\sin^{2}\theta}{4}-\dfrac{J_{1}\sin^{2}\theta}{4}\right]. \nonumber
\end{eqnarray}
Here momentum conservation law is $\textbf{k}_{1}+\textbf{k}_{2}+\textbf{k}_{3}+\textbf{k}_{4}=0$.

Accounting for $\mathcal{H}_3$ and $\mathcal{H}_4$ terms leads to diagrams shown in Fig.~\ref{diagrams}.


\section{Renormalization of anisotropy constant}
\label{AppB}

Oce can see that terms $\mathcal{H}_{0}$, $\mathcal{H}_{1}$ and $\mathcal{H}_{2}$ contain the renormalized single-ion anisotropy constant $\tilde{D}$ instead of $D$. Here we describe this fact.

The renormalization  $\tilde{D}$ of single-ion anisotropy constant arises from the bosonic commutation relations \cite{Lindgard_1974, Lindgard_1976, Rastelli_1979, Balucani_1979, Balucani_1980, Rezende_1983}. After application of relations~\eqref{locspin} and \eqref{spinresp} three operators part $\mathcal{H}_3$ and four operators part $\mathcal{H}_4$ of the Hamiltonian do not appear in the normal form. Transformation of the Hamiltonian to a normal form with the use of bosonic commutation relations leads to the appearance of certain corrections. To illustrate this let's consider the single-ion anisotropy term after applying relation~\eqref{locspin}:
\begin{equation}
D\sum_{i}({S_{i}^{z}})^{2}=D\sum_{i}{(S_{i}^{z^\prime}\cos{\theta}-S_{i}^{y^\prime}\exp(i\textbf{k}_{0}\textbf{R}_{i})\sin{\theta}})^{2} \nonumber
\end{equation}
For example, we will consider only first term coming up from square of $S_{i}^{z^\prime}$ by making transformation~\eqref{spinresp},
\begin{eqnarray*}
D\sum_{i}{(S_{i}^{z^\prime}\cos{\theta}})^{2}= D\cos^{2}{\theta}\sum_{i} \left[(S-a^{\dagger}_{i}a_{i})(S-a^{\dagger}_{i}a_{i})\right]\nonumber=\\D\cos^{2}{\theta}\sum_{i}\left[S^{2}-2Sa^{\dagger}_{i}a_{i}\nonumber+a^{\dagger}_{i}a_{i}a^{\dagger}_{i}a_{i}\right]=\\ D\cos^{2}{\theta}\sum_{i}\left[S^{2}-2Sa^{\dagger}_{i}a_{i}+a^{\dagger}_{i}a_{i}\nonumber+a^{\dagger}_{i}a^{\dagger}_{i}a_{i}a_{i}\right]=\\D\cos^{2}{\theta}\sum_{i}\left[S^{2}-2S\left(1-\dfrac{1}{2S}\right)
a^{\dagger}_{i}a_{i}+a^{\dagger}_{i}a^{\dagger}_{i}a_{i}a_{i}\right]=\nonumber\\\nonumber\cos^{2}{\theta}\sum_{i}\left[DS^{2}-2S\tilde{D}a^{\dagger}_{i}a_{i}+Da^{\dagger}_{i}a^{\dagger}_{i}a_{i}a_{i}\right].\nonumber
\end{eqnarray*}
Evidently the bilinear term contains $\tilde{D}$, which arose from the four operators term and bosonic commutation relations~\cite{Kaganov_1987, Lindgard_1974}. In the same way $\tilde{D}$ will appear in  $\mathcal{H}_{1}$ from three operators terms.

\bibliography{Magnons}

\end{document}